\documentclass{article} 

\usepackage{amsmath,amssymb,amsthm,latexsym} 

\usepackage{stmaryrd,wasysym,upgreek,mathrsfs,dsfont} 
\usepackage[english]{babel} 
\usepackage{graphicx,color} 

\newcommand{\bbbone}{{\mathds{1}}}

\newcommand{\be}{\begin{equation}}
\newcommand{\ee}{\end{equation}}
\newcommand{\bea}{\begin{eqnarray}}
\newcommand{\eea}{\end{eqnarray}}

\newcommand{\cE}{{\cal{E}}}

\newcommand{\cT}{{\cal{T}}}
\newcommand{\cP}{{\cal{P}}}
\newcommand{\cF}{{\cal{F}}}

\begin{document} 
 \title{Constructive Field Theory in Zero Dimension}
\author{V. Rivasseau\\
Laboratoire de Physique Th\'eorique, CNRS UMR 8627,\\ 
Universit\'e Paris XI,  F-91405 Orsay Cedex, France}

\maketitle 
\begin{abstract} 
Constructive field theory can be considered as a reorganization of 
perturbation theory in a convergent way.
In this pedagogical note we propose to wander through five different methods
to compute the number of connected graphs of the zero-dimensional $\phi^4$ field theory,
in increasing order of sophistication and power.
\end{abstract} 

\section{Introduction} 

New constructive Bosonic field theory methods have been recently proposed which are based
on applying a canonical forest formula to repackage perturbation theory in a better way. This allows to compute the connected quantities of the theory by the same formula but summed over trees instead of forests\footnote{Constructive Fermionic field theory is easier and  was repackaged in terms of trees much earlier
\cite{Les,FMRT1,AR2}.}. The resulting formulation of the theory is
given by a convergent rather than divergent expansion. In short 
this is because there are much less trees than graphs, but they still
capture the vital physical information, which is connectedness.

Combining such a forest formula with the intermediate field method
leads to a convenient resummation of $\phi^4$ perturbation theory.
The main advantage of this formalism over previous cluster and Mayer expansions
is that connected functions are captured by a single formula, and e.g. a Borel summability theorem 
for matrix $\phi^4$ models can be obtained which scales correctly with the size of the matrix 
\cite{R1}. The resulting 
method applies to ordinary
quantum field theory on commutative space as well \cite{MR1}. However in this
point of view the connected functions still involve functional integrals over many replicas of
the intermediate field.

An other even more recent constructive point of view \cite{GMR} is that a 
quantum Euclidean Bosonic field theory is a particular positive scalar 
product on a universal vector space spanned by "marked trees".
That scalar product is obtained by applying a tree or forest formula to the ordinary perturbative expansion 
of the QFT model under consideration. That formula itself is model-independent and reorganizes perturbation theory 
differently,  by breaking Feynman amplitudes into pieces and putting these pieces into boxes labeled by trees. 

In this point of view, constructive bounds reduce essentially to the positivity of the 
universal Hamiltonian operator. The vacuum is the trivial tree and  
the correlation functions are given by  ``vacuum expectation values" of the resolvent
of that combinatoric Hamiltonian operator.
Model-dependent details such as space-time dimension, 
interactions and propagators enter the definition of the matrix elements 
of this scalar product. These matrix elements are just finite sums 
of finite dimensional Feynman integrals.

We were urged to explain these new ideas in a pedagogical way.
This is what we do in this short note on the simplest possible example,
namely the connected graphs of the zero dimensional $\phi^4$ theory. This theory corresponds to an
ordinary integral on a single variable, and we hope that following the different constructive steps
on this simple example will expose the core ideas better. The main picture that emerges is that the essence of constructive theory is about using cleverly trees
and replicas.


\medskip\noindent{\bf Acknowledgments}

We thank A. Abdesselam and P. Leroux for the organization of a very stimulating workshop on combinatorics and physics, during which B. Faris asked for such a pedagogical note. We also thank
the IHES where these ideas were presented as part of a course at the invitation of A. Connes.
We also thank Jacques Magnen for a lifelong collaboration
on the topic of constructive theory, only very slightly touched upon in this pedagogical note.
Finally we thank Matteo Smerlak for his critical reading of this manuscript.

\section{The Forest Formula}

Consider $n$ points; the set of pairs $P_n$ of such points which has
$n(n-1)/2$ elements $\ell = (i,j)$ for $1\le i < j \le n$, and a smooth function $f$
of $n(n-1)/2$ variables $x_\ell$, $\ell \in \cP_n$. Noting $\partial_\ell$ 
for $\frac{\partial}{\partial x_\ell}$, the standard canonical forest formula is \cite{BK,AR1}

\be\label{treeformul1}
f(1,\dots ,1)
= \sum_{\cF}  \big[ \prod_{\ell\in \cF}   
\int_0^1 dw_\ell   \big]� \big( [ \prod_{\ell\in \cF} \partial_\ell ] f 
\big)  [ x^\cF_\ell (\{ w_{\ell'}\} ) ]
\ee
where 
\begin{itemize}

\item the sum over $\cF$ is over forests over the $n$ vertices, including the empty one 

\item $x^\cF_\ell (\{ w_{\ell'}\} )$ is the infimum of the $w_{\ell'}$ for $\ell'$
in the unique path from $i$ to $j$ in $\cF$, where $\ell = (i,j)$. If there is no such path,
$x^\cF_\ell (\{ w_{\ell'}\} ) = 0$ by definition.

\item The symmetric $n$ by $n$ matrix $X^\cF (\{w\})$ defined
by $X^\cF_{ii} = 1$ and $X^\cF_{ij} =x^\cF_{ij} (\{ w_{\ell'}\} ) $ 
for $1\le i < j \le n$ is positive.

\end{itemize}

A particular variant of this formula (\ref{treeformul1})  is in fact
better suited to direct application to the parametric representation 
of Feynman amplitudes. It consists in changing variables $x \to 1-x$
and rescaling to $[0,1] \to [0,\infty]$ of the range of the variables.
One gets that if $f$ is smooth with well defined 
limits for any combination of $x_\ell$ tending to $\infty$,

\be\label{treeformul2}
f(0, \dots, 0) = \sum_{\cF}   \big[ \prod_{\ell\in \cF}   
\int_0^\infty  ds_\ell   \big]� \big( [ \prod_{\ell\in \cF} - \partial_\ell ] f 
\big)  [ x^\cF_\ell (\{ s_{\ell'}\} ) ]
\ee
where 
\begin{itemize}

\item the sum over $\cF$ is like above,

\item $x^\cF_\ell (\{ s_{\ell'}\} )$ is the \emph{supremum} of the $s_{\ell'}$ for $\ell'$
in the unique path from $i$ to $j$ in $\cF$, where $\ell = (i,j)$. If there is no such path,
$x^\cF_\ell (\{ s_{\ell'}\} ) = \infty$ by definition. This is because the change of variables
exchanged $\inf$ and $\sup$.
\end{itemize}

To distinguish these two formulas we call $w$ the parameters of the first one (like ``weakening")
since the  formula involves infima, and $s$ the parameters of the second one (like ``strengthening" or ``supremum") since the formula involves suprema.

Scale analysis is the key to renormalization, and scales can be conveniently defined in quantum field
theory through the parametric repesentation of the propagator $H^{-1} = \int_0^{\infty} e^{- \alpha H}
d \alpha$. The
strengthening formula (\ref{treeformul2}) rather than 
the weakening formula is therefore particularly adapted to scale analysis and renormalization.

Finally notice that various extensions of these formulas should be useful to study
furter the theory. Calling {\it endotree} 
a monocyclic connected graph, endotree formulas coud be better
adapted  to the study of vacuum graphs, $p$-particle irreducible 
formulas could be more adapted to the study of particles and so on.
The general theory of such formulas is given in \cite{A}.

\section{Borel summability}
Borel summability of a power series $\sum_n a_n \lambda^n $ 
means existence of a function $f$ with two properties \cite{Sok}:

\begin{itemize}

\item  Analyticity in a disk tangent  to the imaginary axis at the origin

\item plus remainder estimates uniform in that disk:   

\bea  \vert f(\lambda) -  \sum_{n=0}^{N}  a_n \lambda^n \vert   \le
K^N \; N!  \; \vert \lambda \vert^{N+1} \; \; .
\eea
\end{itemize}

Given any power series $\sum_n a_n \lambda^n $,
there is at most one such function $f$. When there is one, 
it is called the Borel sum, and it can be computed from the series to arbitrary accuracy.

Therefore Borel summability is a perfect substitute for ordinary analyticity 
when a function is expanded at a point on the frontier of its analyticity domain.
Borel summability is just a rigidity which plays the same role than analyticity: it 
selects a {\it unique} map between a certain class of functions and a certain class
of power series. Within that class, all the information about the function
is therefore captured in the much more compact list of its Taylor coefficients.

Very early, both functional integrals and the Feynman perturbative series were introduced  to study quantum field theory, but it was realized that the corresponding power series were
generically divergent. When a link between both approaches can be established 
it is usually through Borel summability or a variant thereof. This is why
Borel summability is important for quantum field theory.

\section{$\Phi^4$ constructive theory in zero dimension}

In this section we propose to test the evolution of ideas in constructive theory 
on the simple example of a single-variable ordinary integral which represents the
$\phi^4$ field theory in zero dimensions.

The normalization or partition function of that theory is the ordinary integral
\be
F(\lambda)= \int_{-\infty}^{+\infty} e^{- \lambda x^4 - x^2/2}  \frac{dx}{\sqrt{2 \pi}}  .
\ee
It is obvious that $F(\lambda)$ is well defined for $Re \lambda \ge 0$ with $\vert F(\lambda ) \vert \le 1$, and analytic
for $Re \lambda >0$. It has a Taylor series at the origin $F \simeq \sum_n a_n (-\lambda)^n $ with $a_n = (4n)!!/n! $.
Using a Taylor expansion with integral remainder it is also very easy to prove that $F$ is Borel summable.
But in physics we are interested in computing connected quantities, hence in
the function $G(\lambda) = \log  F(\lambda)$. We  can  therefore formulate the simplest of all Bosonic 
constructive field theory problems as:

\noindent
{\bf Problem}  {\it How to compute $G(\lambda)$ and prove its  Borel summability in the most explicit and efficient manner?}

We shall review how different methods of increasing sophistication answer this question.

\begin{itemize}
\item
Composition of series (XIXth century)
 
\item
A la Feynman (1950)

\item
``Classical Constructive", \`a la Glimm-Jaffe-Spencer (1970's-2000's)

\item
With loop vertices (2007)

\item With tree vector space (2008),
\end{itemize}

\subsection{Composition of series}

The first remark is that we know the explicit power series for $F$, and it starts with 1, so that $F=1+H$;
we know the explicit series for $\log (1+x)$ so we can substitute $H$ for $x$, reexpand and we get
an explicit formula for the coefficient $b_n$ of the Taylor series of $G$.

\bea  F &=& 1 + H ,  \ H = \sum_{p \ge 1}  a_p  (-\lambda)^p , \  a_p = \frac{(4p)!!}{p!}  \  \nonumber \\ \nonumber
\log (1+x) &=&  \sum_{n=1}^{\infty}  (-1)^{n+1}  \frac{x^n}{n} ,  \\ 
 G  &=&  \sum_{n=1}^{\infty}  (-1)^{n+1}  \frac{H(\lambda)^n}{n}   =  
\sum_{k \ge 1} b_k (- \lambda)^k , \  \nonumber \\ 
b_k &=&  \sum_{n=1}^k   \frac{ (-1)^{n+1} }{n}
\sum_{p_1, .. , p_n \ge 1 \atop p_1 + ... +p_n =k}  
 \prod_j   \frac{(4p_j)!!}{p_j!}\; .
\eea

To summarize, this method leads to an explicit formula for $b_k$, but of little use. 
Borel summability of $G$ is unclear. Even the sign of $b_k$ is unclear.

\subsection{A la Feynman}

Feynman understood that $a_n$ can be represented as a sum of terms associated to drawings, 
the famous Feynman graphs. But the real mathematical power of this idea is that it allows a quick computation of logarithms: they are simply given by the same sum but over connected drawings!

The theory of combinatoric species is a rich mathematical orchestration 
of this intuition, see \cite{BLL}. A species is roughly a structure on finite sets of ``points", and combinatorics consists in counting the number $a_n$ of elements in that species on $n$ points.
A fundamental tool to this effect is the generating power series 
of the species which is $\sum_n \frac{a_n}{n!} \lambda^n$.
Let us say that a species {\it proliferate} if
its generating power series has zero radius of 
convergence. A big problem of the current formulation of perturbative quantum field theory
is that the species of ordinary Feynman graphs
proliferate, whereas trees don't. The main problem of constructive theory is therefore to replace
Feynman graphs by trees.

In our case the drawings are the Wick contractions of $\phi^4$ vacuum graphs, that is graphs on $n$
vertices with coordination 4 at each vertex (loops, called
{\it tadpoles} by physicists, being allowed).

\be F = 1 + H ,  \ H = \sum_{p \ge 1}  a_p  (-\lambda)^p , \  a_p =  \frac{1}{p!} \#\{ {\rm vacuum\  graphs\  on} \  p \ {\rm vertices}\}
\ee

\bea G  &=&  \sum_{k=1}^{\infty}  (-\lambda)^{k} b_k,   b_k = \frac{1}{k!} \#\{ {\rm vacuum\ 
connected\  graphs\  on} \  k \ {\rm vertices}\}
\nonumber
\eea
\medskip
We can easily compute in this way that $b_1 = 3$,  $b_2 = 48 $, $b_3 = 1584$...

Usually in the quantum field theory literature there are painful discussions
on what is a Feynman graph and what is its combinatoric weight, or ``symmetry factor".
This is important to make the shortest possible list of independent Feynman amplitudes that one
has to compute in practice. But conceptually it is much better to consider a graph
as a set of ``Wick contractions", that is a set of pairing of fields, so that no ``symmetry factors" are ever discussed\footnote{Fields correspond to half-lines, also called
flags in the mathematics literature and the physicists point of view that flags, not lines, are the fundamental elements in graph theory is slowly making its way in the mathematics literature \cite{Kauf}.}.

\medskip

Borel summability remains unclear. But as a first fruit of the idea of Feynman graphs
clearly, we now see explictely that $b_k \ge 0 $: we know that $G$ has an alternate power series.

\subsection{Classical Constructive}

The standard method in Bosonic constructive field theory is to first break up the functional integrals
over a discretization of space-time, then test the couplings between the corresponding functional integrals 
(cluster expansion), which results in the theory being written as a polymer gas with hardcore constraints.
For that gas to be dilute at small coupling, the normalization of the free functional integrals must be factored out.
Finally the connected functions are computed by expanding away the hardcore constraint through a so-called
Mayer expansion \cite{GJ,Riv1,Br,AR}.

In the zero dimensional case there is no need to discretize the single point of space-time, hence it seems  that the first step, namely the cluster expansion is trivial. This is correct except for the fact that what remains from this step is to factorize the ``free functional" integral, so a single first-order Taylor expansion
with remainder around $\lambda=0$ corresponds to the zero-dimensional cluster expansion. But after that
the Mayer expansion is completely non-trivial and very instructive, 
because a single point has indeed hardcore constraint with itself!

The first step (cluster expansion) is therefore:
\be  F = 1 + H ,  \ H = - \lambda \int_0^1  dt  \int_{- \infty}^{+ \infty}   x^4 e^{- \lambda t x^4 - x^2/2}  \frac{dx}{\sqrt{2 \pi}}\; .
\ee
The more interesting Mayer expansion
in this case consists in introducing many copies or ``replicas" of $H$:
\be \forall i\quad H= H_i  = - \lambda \int_0^1  dt  \int_{- \infty}^{+ \infty}   x_i^4 e^{- \lambda t x_i^4 - x_i^2/2}  \frac{dx_i}{\sqrt{2 \pi}}\;  , 
\ee
Defining $\epsilon_{ij} = 0 \  \forall i, j $ we can write the apparently stupid formula
\be  F = 1 + H =  \sum_{n=0}^{\infty}  \prod_{i=1}^n  H_i (\lambda)  \prod_{1 \le i < j \le n} \epsilon_{ij}.
\ee
But defining $\eta_{ij} = -1$, $\epsilon_{ij} = 1 + \eta_{ij} = 1+x_{ij} \eta_{ij} \vert_{x_{ij} = 1}$ and applying the forest formula leads to
\be F = \sum_{n=0}^{\infty}   \frac{1}{n!}\sum_{\cF}  \prod_{i=1}^n H_i (\lambda)   \bigg\{ \prod_{\ell\in \cF}   
\big[ \int_0^1 dw_\ell \big]    \eta_\ell \bigg\} \prod_{\ell \not\in \cF} \big[ 1 +  \eta_{\ell}  x^\cF_\ell (\{ w\}) \big]\, ,
\ee
which allows easily to take the logarithm
\be G = \sum_{n=1}^{\infty}   \frac{1}{n!}  \sum_{\cT}  \prod_{i=1}^n H_i (\lambda)   \bigg\{ \prod_{\ell\in \cT}   
\big[ \int_0^1 dw_\ell \big]    \eta_\ell \bigg\} \prod_{\ell \not\in \cT}   \big[1 +  \eta_{\ell} x^\cT_\ell (\{ w\})   \big]\; .
\ee
where the second sum runs over trees!  In this way we obtain that

\begin{itemize}
\item
Convergence is now easy because each $H_i$ contains  a different "copy"  $\int dx_i$ of 
the ``functional integration" (which of course here is an ordinary integration).

\item
Borel summability for $G$ follows now easily from the Borel summability of $H$.
\end{itemize}

A shortcoming is that ``space-time" and functional integrals remains present.
Also the cluster step, suitably generalized in non zero dimension
by Glimm, Jaffe, Spencer and followers, heavily relies on locality,
hence does not seem to have the potential to work on nonstandard space-times
or for nonlocal or matrix-like theories like the Grosse-Wulkenhaar model of  noncommutative theory
\cite{GW} or for the group field theory models of quantum gravity. 

\subsection{Loop Vertices}

The intermediate field representation is a well known trick to represent a quartic interaction in terms
of a cubic one:

\bea F &=&  \int_{- \infty}^{+ \infty}  e^{- \lambda x^4 - x^2/2}  \frac{dx}{\sqrt{2 \pi}}  =   \int_{- \infty}^{+ \infty}  \int_{- \infty}^{+ \infty}  e^{- i \sqrt{2\lambda }\sigma x^2 - x^2/2  - \sigma^2/2}  \frac{dx}{\sqrt{2 \pi}} \frac{d\sigma}{\sqrt{2 \pi}} \nonumber \\
&=&  \int_{- \infty}^{+ \infty}  e^{- \frac{1}{2} \log [ 1 + i \sqrt{8 \lambda} \sigma ] - \sigma^2/2}  \frac{d\sigma}{\sqrt{2 \pi}}
\nonumber \\
&=&  \int_{- \infty}^{+ \infty}  \sum_{n=0}^\infty  \frac{V^n}{n!} d\mu (\sigma)\; , \quad {\rm with} \; \;
 V= - \frac{1}{2} \log [ 1 + i \sqrt{8 \lambda} \sigma ]\; . \label{rig}
\eea

We can introduce again replicas but in a slightly different way.
We duplicate the intermediate field into copies, 
$V^n (\sigma) \to  \prod_{i=1}^n V_i (\sigma_i)$, one associated to each factor $V$, also called a ``loop vertex".
The theory has not changed if we use for all these fields a jointly Gaussian measure
with degenerate covariance, $d\mu (\sigma) \to d\mu_C ( \{\sigma_i \})$, $C_{ij} = 1  = x_{ij}\vert_{x_{ij} = 1}$. Remark that the corresponding measure has no density with respect to the Lebesgue measure
since
\be d\mu_C ( \{\sigma_i \}) =
 \frac{d\sigma_1}{\sqrt{2 \pi}} e^{- \sigma_1^2/2} 
\prod_{i=2}^n  \delta(\sigma_1 - \sigma_i ) d\sigma_i \; . 
\ee
It is not enough known that a delta function {\it is} a Gaussian measure!

This does not change the expectation value of any polynomial, hence by the Weierstrass
theorem it does not change the theory. 
But one can now apply the forest formula to the off-diagonal couplings $C_{ij}$. It gives:
\be F = \sum_{n=0}^{\infty}   \frac{1}{n!}  \sum_{\cF}  \bigg\{ \prod_{\ell\in \cF}   
\big[ \int_0^1 dw_\ell \big]   \bigg\}  {\bf \int}  
 \bigg\{ 
 \prod_{\ell\in \cF}   
\frac{\partial }{\partial \sigma_{i(\ell)} } 
\frac{\partial }{\partial \sigma_{j(\ell)} } 
\prod_{i=1}^n V(\sigma_i)
 \bigg\} 
   d\mu_{C^{\cF}} 
\ee
where $C^{\cF}_{ij}  =  x^\cF_\ell (\{ w\})$ if $i < j$, $C^{\cF}_{ii}  = 1$.

\be G = \sum_{n=1}^{\infty}   \frac{1}{n!}  \sum_{\cT}  \bigg\{ \prod_{\ell\in \cT}   
\big[ \int_0^1 dw_\ell \big]   \bigg\}  {\bf \int}  
 \bigg\{ 
 \prod_{\ell\in \cT}   
\frac{\partial }{\partial \sigma_{i(\ell)} } 
\frac{\partial }{\partial \sigma_{j(\ell)} } 
\prod_{i=1}^n V(\sigma_i)
 \bigg\}  d\mu_{C^{\cT}} ,
\ee
where the second sum runs over trees!  

The main advantage is that the role of propagators and vertices have been exchanged!
The result is a sum over trees on loops, or \emph{cacti}. 

\begin{figure}[!htb]
\centering
\includegraphics[scale=0.5]{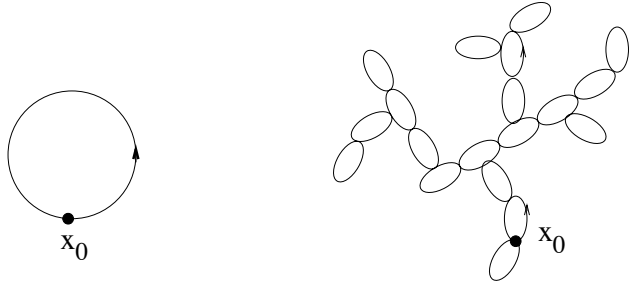}
\caption{Loop vertices and a tree on them, or cactus.}
\label{looptree}
\end{figure}

Since 
\be \frac{\partial^k }{\partial \sigma^k } \log [ 1 + i \sqrt{8 \lambda} \sigma ]  = -(k-1)!  (-i \sqrt{8 \lambda} )^k [  1 + i \sqrt{8 \lambda} \sigma ]^{-k} , \nonumber\ee
the loop vertices involve denominators or ``resolvents"  $[  1 + i \sqrt{8 \lambda} \sigma ]^{-1}$  
rather than $\log$'s. However
there is a little subtlety with the first ``trivial" term in $G$ for $n=1$ which is a single
vertex loop with value $\log [ 1 + i \sqrt{8 \lambda} \sigma ]$. To transform it also into an 
expression with denominators one should perform a single Taylor expansion step
and integrate the $\sigma$ field through integration by parts:
\bea \int d \mu (\sigma)  \log [ 1 + i \sqrt{8 \lambda} \sigma ] &=& \int d \mu (\sigma)
\int_0^1 dt  \frac {i \sqrt{8 \lambda} \sigma }
{ 1 + i \sqrt{8 \lambda} t \sigma } \nonumber\\
&=& \int d \mu (\sigma)
\int_0^1 dt  \frac {  8 \lambda t} { [1 + i \sqrt{8 \lambda} t \sigma ]^2 }
\eea

Once $G$ has been rewritten in this way:

\begin{itemize}
\item
Convergence is easy because $\vert [  1 + i \sqrt{8 \lambda} \sigma ]^{-k} \vert  \le 1$,
and because recall that trees do not proliferate.

\item
Borel summability is easy.
\item

This method extends to non commutative field theory and gives correct estimates for matrix-like models.

\end{itemize}

The drawback is that the method is more difficult when the interaction is of higher degree e.g. $\phi^6$,
because more intermediate fields are needed. Moreover
functional integrals over the intermediate fields are still present.

\subsection{Tree QFT}

This last method no longer requires functional integral at all! It is in a way
the closest to Feynman graphs, hence looks at first sight like a 
step backwards in constructive theory. It starts exactly like
constructive Fermionic theory.

Within a given quantum field model, the forest formula indeed associates 
a natural amplitude $A_T$  to a tree $T$. It is the sum of all contributions 
associated to that tree when one applies the tree formula to the $n$-th 
order of perturbation theory of {\it that} model.

In our zero dimensional case it means that we start with the usual
\emph{formal} perturbation theory 
\bea F &=&  \int_{- \infty}^{+ \infty}  e^{- \lambda x^4 - x^2/2}  \frac{dx}{\sqrt{2 \pi}} = \int  \sum_{n=0}^\infty  \frac{V^n}{n!} d\mu (x)\; , 
 \eea
where $ V=(- \lambda x^4)$ and $d\mu (x)$ is the Gaussian measure of covariance 1.
Beware that this interchange of sums and integrals is not licit, contrary to (\ref{rig});
That's why the result, namely ordinary perturbation theory diverges!  Nevertheless
we shall use this perturbation series in a heuristic way, namely we shall repackage 
according to a forest formula and use the pieces so obtained to build a 
(semidefinite) scalar product on a vector space generated by marked trees. 
This will allow still another rigorous constructive expansion of the function $G$.

Then we introduce replicas again but on the ordinary vertices:
\bea F &=&   \sum_{n=0}^{\infty}\int \frac{1}{n!}\prod_{i=1}^n(- \lambda x_i^4) d\mu(\{x_i\})
\eea
where $d\mu$ is the degenerate measure with covariance $C_{ij} = 1 \; \forall i, j $.

Applying the tree formula to that covariance gives in the same vein than before
\be F = \sum_{n=0}^{\infty}   \frac{1}{n!}  \sum_{\cF}  \bigg\{ \prod_{\ell\in \cF}   
\big[ \int_0^1 dw_\ell \big]   \bigg\}  {\bf \int}  
 \bigg\{ 
 \prod_{\ell\in \cF}   
\frac{\partial }{\partial x_{i(\ell)} } 
\frac{\partial }{\partial x_{j(\ell)} } 
\prod_{i=1}^n (- \lambda x_i^4)
 \bigg\} d\mu_{C^{\cF}} 
\ee
where $C^{\cF}_{ij}  =  x^\cF_\ell (\{ w\})$ if $i < j$, $C^{\cF}_{ii}  = 1$, so that {\it formally}
\be G = \sum_{n=1}^{\infty}   \frac{1}{n!}  \sum_{\cT}  \bigg\{ \prod_{\ell\in \cT}   
\big[ \int_0^1 dw_\ell \big]   \bigg\}  {\bf \int}  
 \bigg\{ 
 \prod_{\ell\in \cT}   
\frac{\partial }{\partial x_{i(\ell)} } 
\frac{\partial }{\partial x_{j(\ell)} } 
\prod_{i=1}^n (- \lambda x_i^4)
 \bigg\} d\mu_{C^{\cT}} 
\ee

The zero-dimensional $\phi^4$ tree amplitude for a tree $\cT$ is therefore nothing but
\be  A_{\cT} = \bigg\{ \prod_{\ell\in \cT}   
\big[ \int_0^1 dw_\ell \big]   \bigg\}  {\bf \int}  
 \bigg\{ 
 \prod_{\ell\in \cT}   
\frac{\partial }{\partial x_{i(\ell)} } 
\frac{\partial }{\partial x_{j(\ell)} } 
\prod_{i=1}^n (- \lambda x_i^4)
 \bigg\} d\mu_{C^{\cT}} .
\ee
Remark that it is zero for trees with degree more than four at a vertex.

It seems little has been achieved at the constructive level by rewriting
Feynman graphs simply in terms of an underlying tree, like in Fermionic theories.
But there is a hidden {\it convexity} in $A_{\cT}$ which hides
the non-perturbative stability of the underlying theory.

It has indeed be shown in \cite{GMR} that one can construct a scalar product over 
an abstract infinite dimensional vector space ${\cal E}$ with a spanning basis
$e_T$ labeled by marked trees, which are trees with a mark
on a particular leaf (ie vertex of degree 1). 
On ${\cal E}$ there is a natural external gluing operation which 
sends $(T,T')$ onto the (unmarked) tree $T\star T'$
by gluing the two marks.

This operation induces a natural (semi-definite) scalar product and a natural
$\phi^4$ Hamiltonian operator $H$. The scalar product $<e_T, e_{T'}> $ is 
simply $A_{T \star T'}$. Roughly speaking
the Hamiltonian operator glues all the trees hanging to a single vertex with two marks.
It is a Hermitian negative operator on the Hilbert space which is the completion of 
$\cE$ for the scalar product above.

For the $\phi^4$ theory one can always restrict to trees with degrees at most 4
at each vertex. 

The constructive expression for 
eg the connected two point function $$G_2 = \frac{1}{F}
 \int_{- \infty}^{+ \infty} x^2 e^{- \lambda x^4 - x^2/2}  \frac{dx}{\sqrt{2 \pi}}$$ 
computed with this method is
\be G_2  = < e_{T_0}, \frac {1}{1 - H} e_{T_0}>\;\; .
\ee
where $T_0$ is the trivial tree with a single line.
It is well defined at non-perturbative level because $H$ is Hermitian negative.

This approach does not require any functional integral, since $\cE$ is spanned by finite order trees
and its scalar product depends only on finite dimensional perturbative computations. It seems
the most promising way to study quantum field theory in the future, including
in non-integer dimensions. However Borel summability is not completely obvious in this expression
hence more work is needed.

Also notice that the expressions of the functions $F$ and $G$ within this method 
should require a slight extension of \cite{GMR}. The tree formulas (\ref{treeformul1})-(\ref{treeformul2})
should be pushed a single Taylor step further to {\it endotree} formulas, see above. 
Vacuum connected graphs should 
be distributed as a sum over endotrees rather than tree, and the space $\cE$
should be enlarged accordingly. The corresponding formalism 
together with the multiscale version of this approach are to be developed.
Still we think this approach is the most general and promising one for the future
of constructive theory, since it is the most abstract and general. Neither functional integrals nor
a space time plays a central role, which makes the theory appealing for situations such as
non integer dimensions or the future theory of quantum gravity.

\end{document}